\documentclass[twocolumn]{article}
\usepackage[T1]{fontenc}
\usepackage[latin1]{inputenc}

\input{tcilatex}
\begin{document}

\title{A Generator of Protein Folding Kinetics States for the
Diffusion-Collisi Model}

\author{Zlatko K. Vasilkoski and David L. Weaver}

\date{March 18, 2000}
\maketitle

{\centering \emph{Molecular Modeling Laboratory,}\par}
{\centering \emph{Department of Physics,}\par}
{\centering \emph{Tufts University,}\par}
{\centering \emph{Medford Massachusetts 02155}\par}
{\centering \emph{Corresponddence: dweaver@tufts.edu }\par}

\begin{quote}
\bigskip
\end{quote}

\begin{abstract}
Two separate algorithms for calculating the intermediate states, using
cellular automata and the initial conditions in the rate matrix for the
diffusion-collision model are introduced. They enable easy and fast
calculations of the folding probabilities of the intermediate states, even
for a very large number of microdomains.
\end{abstract}

\begin{center}
Introduction
\end{center}

In recent years, many theoretical and experimental studies have focused on
the problem of describing the mechanism of protein folding. The goal is to
develop a model that predicts protein folding rates and their dependence on
factors such as temperature, amino acid sequences and so on. There are two
aspects to the prediction problem: one is predicting the native structure of
a protein from its sequence, which is thermodynamic in character; the other
concerns the mechanism by which denatured proteins fold to their native
conformation, and is dynamic in character. The dynamical aspects of folding
are often formulated taking into account Levinthal's paradox \cite{1} that a
random search of all possible structures will result in a time longer than
the age of the universe.

\medskip A model that gives satisfactory predictions regarding the dynamical
aspects of the protein folding is the diffusion-collision model of Karplus
and Weaver \cite{2}, \cite{4} (see Burton, Myers and Oas \cite{4} for a
recent experimental test of the model). This model considers the protein to
be made of secondary structure elements - microdomains, each short enough,
for a rapid conformational search so that Levinthal's paradox is avoided.
Microdomain - microdomain folding is considered to occur as diffusion
through solution, with some collisions between microdomains leading to
smaller and then larger structures, until the native conformation is
reached. The randomness of the diffusion process indicates a very important
characteristic of the model: that the folding process may involve many
possible paths, not just one pathway, leading to the folded state.
Furthermore, the model can calculate the probabilities of kinetic
intermediate states at any moment in time.

\bigskip

\begin{center}
DESCRIPTION OF THE DIFFUSION-COLLISION MODEL
\end{center}

- The diffusion-collision model has the following properties:

- Presence of microdomains;

- Transient secondary structure is formed before tertiary structure;

- Transient accumulation of kinetic intermediates;

- Existence of folding pathways;

- Possible existence of non-native intermediates from non-native collisions;

- Solvent viscosity dependence of folding rates;

- Folding rates and favored pathways dependent on properties of microdomains;

- The microdomains move diffusively under the influence of internal and
random external forces, and microdomain-microdomain collisions occur. The
dynamics of folding is simulated by a set of diffusion equations that
describe the motion of the microdomains in aqueous solution, and by boundary
conditions that provide for the microdomains collision and possible
coalescence. The diffusion-collision dynamics is represented as a network of
steps, each containing a microdomain pair interaction, in which the rate of
coalescence depends on the physical properties of the microdomains. The
rates can be analytically expressed in terms of the physical parameters of
the system.

Let us consider the following analytical model to calculate the folding rate
of two connected microdomains, which is the elementary step in the
diffusion-collision model. Consider two connected microdomains A and B that
coalesce into AB.

\begin{center}
\begin{equation}
A+B->AB
\end{equation}
\end{center}

The dynamical behavior of the microdomains is modeled by a diffusion
equation. Since we have a system of two microdomains, the equations are
coupled \cite{3}. The relative motion diffusion equation is \cite{3}:

\begin{center}
\begin{equation}
\frac \partial {\partial t}\binom{\rho _1}{\rho _2}=D\nabla ^2\binom{\rho _1%
}{\rho _2}+\left( \QATOP{-\lambda _1}{\lambda _1}\QATOP{\lambda _2}{-\lambda
_2}\right) \binom{\rho _1}{\rho _2}
\end{equation}
\end{center}

where $\rho $ is a 2 element vector, $\rho _1$ being a probability density
for both microdomains folded, and $\rho _2$ the probability density for all
other possibilities. D is the relative diffusion constant, and the rate
constants are $\lambda _1$ - from both folded state to all others, and $%
\lambda _2$ the rate for the reverse process. Equation 2 couples
microdomain-microdomain relative diffusion with the two-state
folding-unfolding process carried out in solution by the microdomains. The
connecting chain between them limits the diffusion space for microdomain -
microdomain relative motion. An idealization is made that microdomains are
spheres connected by a polypeptide chain considered to be a flexible
featureless string. The collision and coalescence of the microdomains are
governed by the boundary conditions for Equation 2. The inner boundary is
the closest approach spherical shell, in terms of van der Waals envelopes of
the microdomains. The closest approach distance of two microdomains is the
sum of their radii $R_{min.}$ The other constraint on the diffusion space is
the maximal radial separation $R_{max}$between the microdomains, determined
by the length of the string between A and B. So we have:

\begin{center}
\begin{equation}
R_{\min }=R_A+R_B
\end{equation}
\strut 
\[
R_{\max }=R_A+R_B+\text{shortest intervening chain lenght} 
\]
\end{center}

The boundary conditions on the probability density are specified as: 
\begin{equation}
\frac{\partial \rho _{1,2}}{\partial r}|_{R\max }=0
\end{equation}

which means that the microdomains cannot get further away from one another
than $R_{max}$and the condition:

\begin{equation}
\frac{\partial \rho _2}{\partial r}|_{R\min }=0
\end{equation}

meaning that the unfolded microdomains can not get closer to one another
than $R_{min.}$Finally: 
\begin{equation}
\rho _1|_{R\min }=0
\end{equation}

indicating that the both states folded probability density r1 is zero at the
inner boundary, meaning that coalescence takes place.

The forward (folding) rate of coalescence of two microdomains to form a bond
is taken to be $k_f=1/\tau _f$ during intermolecular diffusion, where $\tau
_f$ is the folding time, and has the following general form \cite{3}: 
\begin{equation}
\tau _f=\frac{l^2}D+\frac{L\nabla V\left( 1-\beta \right) }{\beta DA}
\end{equation}

Following Ref. 3, $D$ is the relative diffusion coefficient,$\nabla V$ is
the volume available for diffusion of each microdomain pair,$A$ is their
relative target surface area for collision,$\beta $ the probability that the
two microdomains are in a folded state, when they collide, so there is no
barrier to coalescence.$L$ and $l$ are geometrical parameters that satisfy
the boundary conditions for the diffusion equation in three dimensions.$L$
has units of length and the value: 
\begin{equation}
\frac 1L=\frac 1{R_{\min }}+\alpha \frac{\alpha R_{\max }\tanh \left[ \alpha
\left( R_{\max }-R_{\min }\right) \right] -1}{\alpha R_{\max }\tanh \left[
\alpha \left( R_{\max }-R_{\min }\right) \right] }
\end{equation}

where $\alpha \equiv \left( \left( \lambda _1+\lambda _2\right) /D\right)
^{1/2}$

For the backward, unfolding rate $k_b=1/\tau _b$, the unfolding time $\tau
_b $, has the following form \cite{3}: 
\begin{equation}
\tau _b=\mathcal{V}^{-1}e^{f\frac{A_{AB}}{k_bT}}
\end{equation}

The actual contact surface area $A_{AB}$ between microdomains $A$ and $B$,
is used as well as $f$, the free energy change per unit area between the
microdomains involved in the bond. The dissociation rate in the absence of
an energy barrier is given by $\mathcal{V}$, $k_b$ is the Boltzmann
constant, and $T$ is the absolute temperature.

Defining the folding and the unfolding rates in this way, for each
two-microdomain process, multi-microdomain protein folding can be treated as
a set of two-microdomain interactions, between all the possible pair
combinations of interactions. The process is continuous in time and the
second order diffusion partial differential equation, reduces \cite{3} to a
system of linear first order (in time) equations represented by the
following vector equation with elements: 
\begin{equation}
\frac{dp_i}{dt}=\stackunder{i=1}{\stackrel{m}{\sum }}K_{ij}p_j
\end{equation}

Here $p_j$ is the probability of a kinetic intermediate state, and would
correspond to the concentration of a substance if the diffusion equation
described the time varying difference in concentration between several
adjacent, spatially discrete regions. The elements $R_{ij}$of the rate
matrix $R$ are determined from the folding and unfolding rates, $k_f$ and $%
k_b$. If $R$can be diagonalized 
\begin{equation}
R=S\Lambda S^{-1}
\end{equation}

then the vector equation can be solved, and by standard linear algebra
procedures finding first the eigenvectors and eigenvalue matrix $\Lambda $
of the rate matrix $R$, the probabilities $p_i$ can be obtained as
exponential functions of time: 
\begin{equation}
p\left( t\right) =p\left( 0\right) Se^{\Lambda t}S^{-1}
\end{equation}

\bigskip

\begin{center}
EXAMPLE OF DIFFUSION-COLLISION MODEL CALCULATIONS
\end{center}

Let us consider a simple protein chain, made of three microdomains as shown
in Figure 1, with the microdomain properties given in Table I.\smallskip
\medskip

\begin{center}

$\stackunder{\text{{\sf The possible pairings for this simple protein chain
are AB, AC and BC.}}}{\stackunder{\text{{\sf A representation of a
3-microdomain unfolded protein chain. }}}{\stackrel{\text{{\sf Figure
1.}}}{\FRAME{itbpF}{141.1875pt}{29pt}{-0.75pt}{}{}{Figure }
{\includegraphics{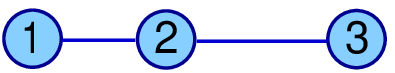}}}}}
\medskip $

$\medskip $
\end{center}

The actual pairings that are possible for this three microdomain protein
are: AB, AC and BC. According to the diffusion-collision model, we have the
schematic description of the possible states shown in Table II.

Here we have 8 possible states, the initial unfolded state \#1, all 6
possible kinetic intermediate states, and the final folded state \#8. The
states are associated with the coalescence of the corresponding pairings
that are indicated in Table II by the digit 1. A schematic view is shown in
Figure 2. Based on the initial data, Table III of transitions, states, bonds
and parameters can be obtained. For larger n-pairings between the
microdomains, obtaining the data in Table III is the most difficult part of
the diffusion-collision calculation. The data from Table III goes into the
calculations of folding and unfolding rates, $k_f$ and $k_b$. Finally, the
rate equation (Eq. 10) is solved to get $p_i$, the probabilities of the
states as a function of time.

Several things can be generalized and noted for $n$ pairings between the
microdomains in a folding protein. First, it can be noted that the number of
states is $2^n$, since we consider two microdomain interaction at a time,
the number of ways in which a population of n elements can be divided into
two sub populations is $2^n$. So we have:

- $2^n$ number of states (8 for $n=3$)

- $n!$ number of independent pathways (6 for $n=3$)

- $n2^{n-1}$ number of transitions (12 for $n=3$)

For a larger number of microdomains and pairings between the microdomains,
calculating the possible states, independent pathways, and all of the
transitions can be a very tedious task. As can be seen from Table IV, the
number of states, independent pathways, and the number of transitions,
increases quickly with increasing n, the number of pairings between the
microdomains of a protein. As n increases, the steepest increase is in the
number of independent pathways, which grows as a factorial. It is
interesting to note that this rapid increase in the number of possible
native pathways may lend evolutionary stability to protein native
structures, since they can be reached via multiple pathways, and blocking of
one or more routes will probably not affect the folded state. From the point
of view of actual calculations, these large numbers create a limiting
problem, finding the actual states and pathways, and more specifically
calculating $R_{max}$, which is a practical problem in a situation when
there are a large number of pathways, corresponding to different
configurations of microdomain interactions.

The two algorithms described below are used to speed up and simplify the
diffusion-collision calculations for a large number of microdomains.

\begin{center}
ALGORITHM FOR OBTAINING DIFFUSION-COLLISION MODEL STATES
\end{center}

As mentioned before, for a given number of pairings $n$, the number of
states is $2^n$. For actual calculations we need to distinguish all of those
states and indicate to which coalesced pairings they correspond. As can be
seen from Table II, we identify these states with numbers, ranging from 1 to 
$2n$. The correspondence with the coalesced pairings is obtained by
numbering the states with binary numbers, where the digit 1 at the
appropriate position underneath the pairing, indicates that coalescence of
that pairing occurs (see Table II). For example, state 1, the unfolded
state, does not have a digit 1 in its binary representation. State 2
indicates that the pairing AB coalesced, so correspondingly there is a digit
1 underneath the pairing AB. The same numbering is applied to all the
states. The binary notation keeps the information of which pairings
coalesce, and they are related to the decimal numbering of the states by
turning the binary number into decimal plus 1. This is a very useful and
condensed way of numbering all the states, except that for large number of
pairings $n$, it is not easy to write down and number all the states. The
following algorithm does that.

It can be noted that $2^n$ states, grouped by the number of pairs in a
state, are actually the binomial coefficients \cite{5}: 
\begin{equation}
\left( 1+x\right) ^n=1+\binom n1x+\binom n2x^2+...+\binom nnx^n=2^n
\end{equation}
where $x=1$, so for one pair states we have $\binom n1$different states, for
two pair states we have $\binom n2$different states, and so on. A schematic
way of representing the binomial coefficients is the familiar Pascals
triangle where the binomial coefficients in the next row are simply related
by addition, to the values in the previous row. This motivates using
cellular automata to create and represent the pair states (binomial
coefficients).\medskip

Cellular Automata

Cellular automata are discrete space-time dynamical systems, where each cell
has a set of possible values, belonging to a finite field. They reproduce on
a space-time grid, evolving synchronously according to a specific
mathematical rule usually in correlation with the number of cell neighbors
on the grid, the most famous example being the Game of Life\cite{6}.

The simplest example of a cellular automaton would be the one-dimensional
case, where the grid is actually made of segments on an infinite line, and
the dependent variable $q_n^t$takes values 0 and 1. Here $t$ and $n$ are
considered as time and space variables, respectively. The initial data is a
set of zeros and ones, and the time evolution is given by the rule function $%
f_r(q_n^t)$ that enables us to construct the next step. So for any
generation we have: 
\begin{equation}
q_n^{t+1}=f_r(q_n^t)
\end{equation}

Here the subscript $r$ indicates the neighborhood of the rule function, or
on how many spatial neighbors ($2r+1$ in this case) the cell at position $k$
depends: 
\begin{equation}
q_k^{t+1}=f_r(q_{k-r}^t,...,q_k^t,...)
\end{equation}

For example, by taking the simple case of $r=1$ (corresponding to nearest
neighbor correlation) and a rule function defined as: 
\begin{equation}
q_k^{t+1}=(q_{n-1}^t+q_{k+1}^t)\equiv q(\func{mod}2)
\end{equation}

where $q$ is either 1 or 0 and $\equiv $ stands for Congruence (integral
divisibility in the sense $0\equiv 0(\func{mod}2);1\equiv 1(\func{mod}%
2);2\equiv 0(\func{mod}2)$ ), we get the pattern in Table V. As we can see,
the rule of multiplication is that the cell multiplies in the next moment of
time (generation) at position n-1 and n+1. In addition, there is the rule of
annihilation (overcrowding) of cells that happens when a cell has two
neighbors (ones). The middle cell becomes zero from overcrowding since two
new cells are born on the same place. These rules apply to all generations
(time steps). This is just one example of a rule function whose pattern is a
picture of Serpinskys triangle, that is, a fractal formed by deleting the
inside triangle of a larger equilateral triangle. Serpinskys triangle can be
also generated from Pascals (binomial coefficients) triangle by deleting the
even numbers from it. This is the motivation for using a kind of a cellular
automaton with a similar rule function as a way of generating the pair
states (binomial coefficients) for the diffusion-collision model. The
missing feature in the rule for succeeding generations is the binomial
representation of the states that actually keeps the information about the
transitions, independent pathways and the pairings involved. To solve this
problem, an additional feature is added to the cellular automata, namely
putting a binary genetic code in the cells of the one-dimensional cellular
automata. By adding a rule to the evolution of the cellular automata,
involving the transfer of the binary genetic code to the next cell, we will
be able to keep track of kinetic intermediates and get all the information
necessary for our purpose of describing the folding kinetics of the states
for the diffusion-collision model.\medskip

Cellular Kinetics

Different folding states are obtained by evolving one-dimensional cellular
automata with a binary genetic code contained in each of the cells. At every
new generation the binary genetic code is mixed with the parents code from
the previous generation. The rule of mixing is the following: in the next
generation, a 0 from the right is attached to the binary genetic code of the
cell on the $n-1$ position, and a $1$ from the right is attached to the
binary genetic code of the cell on the $n+1$ position. For example, starting
from generation $t$:
\begin{verbatim}
Generation (t):
cell [ 011010 ]
Generation (t+1):
cell [ 011010 <-0 ] cell [ 011010 <-1 ]
\end{verbatim}

The binary genetic code for the first few generations is:
\begin{verbatim}
Gen. (t=0)          [0]
Gen. (t=1)    [0<-0]   [0<-1]
\end{verbatim}

After the first generation, to the initial binary genetic code [0] at $t=0$
and $n$-th position, we have one zero added from the right at $t=1$; $(n-1)$%
-th position obtaining [00] and one $1$ added from the right at $t=1$; $%
(n+1) $-th position obtaining [01].
\begin{verbatim}
Gen. (t=2)    [00<-0]  [00<-1]  [01<-1]
Gen. (t=2)             [01<-0]
\end{verbatim}

From the second generation each cell multiplies to the left and to the right
adding a 0 or 1 respectively to the end of its own binary genetic code,
forming the following cells:
\begin{verbatim}
Gen. (t=3) [000<-0] [000<-1] [001<-1] [011<-1]
Gen. (t=3)          [001<-0] [010<-1]
Gen. (t=3)          [010<-0] [011<-0]
\end{verbatim}

We can see that by applying this rule we get exactly the states we need for
our purpose of describing the diffusion-collision folding kinetics.
Continuing this simple rule will list all the states in binary/decimal form
for fast and easy generation of a large number of pairings. Enumerating all
the possible states in this manner is very convenient for keeping track of
the independent pathways and transitions for diffusion-collision folding
kinetics.\medskip

\begin{center}
ALGORITHM FOR OBTAINING $R_{max}$
\end{center}

\medskip A second important problem in calculations of diffusion-collision
folding kinetics for a large number of pairings $n$, is finding the upper
bound of the diffusion space -$R_{max}$. As mentioned before, $R_{max}$
needs to be found for every transition and is then used to calculate the
folding rate $k_f$ for that transition. The problem becomes complicated even
for a small number of pairings, like $n=5$ or $6$. The different microdomain
conformations make it very hard to find what is the maximal distance between
the pairings that are supposed to coalesce. Below is a schematic
illustration of the problem and the algorithm that is used to solve it.

In the diffusion-collision model, the microdomain structure of a protein is
assumed to be like a number of beads on a string, all having different radii
and distances between one another. In its unfolded state the protein will
look, for example, like the string of 11 beads (microdomains) shown in
Figure 3. To find the maximum distance between coalescing microdomains, for
example beads 2 and 7, we need only to sum up the distances 2-3, 3-4, 4-5,
5-6, 6-7, the diameters of beads 3,4,5 and 6 and the radii of beads 2 and 7.
This will be the maximum size of the (spherical) diffusion space. The
minimum size is the radial distance between their centers, when bead 2 and 7
are in contact. The situation quickly becomes complicated when more beads
(microdomains) start to coalesce. For example, if we have to find again the
maximum distance between beads 2 and 7, but beads 3 and 9 have already
coalesced, then we need to worry about additional paths between beads 2 and
7. We will then have several different ways of calculating the distances.
For example, from 2 to 7 we can go through 2-3, 9-4-5-6-7, or 2-3, 9-8-7, as
illustrated in Figure 4. In this case, we need to find the shortest of the
possible paths between 2 and 7. Depending on the distances and radii of the
beads, any of the available ways for a given conformation can be the
shortest. As we have more and more beads (microdomains) coalescing, the
situation soon gets very complicated. In order to analyze the different
confirmations, we need first to have a way of keeping track of them. That
can be done by using matrices. The N-microdomain protein structure is kept
in the adjacency matrix $(AM)$. It is a N x N matrix where the nonzero
elements indicate the adjacent neighbors. For our example of $N=11$
microdomains, the adjacency matrix $(AM)$ for the unfolded state is given in
Table VI. Coalescence between microdomains 3 and 9 changes the adjacency
matrix $(AM)$, so that 3 is now a neighbor to 10 and 8, and 9 is a neighbor
to 2 and 4, which is indicated in the new adjacency matrix in Table VII, as
new nonzero elements.

In order to find $R_{max}$ between two beads i and j, once we have a way of
tracking the different conformations, we need to consider all the possible
ways of getting from i to j, for a given conformation of the other beads.
Multiplying the adjacency matrix $AM$ with itself does that. The power $n$
of $AM^n$ will correspond to the steps between the beads. The new nonzero
elements in the product matrix contain all of the possible routes to get
from the initial bead i to n. In order to get to j we need to repeat the
procedure at most $AM^{(j-i)}$ times. This will cover all of the possible
steps (distances) between the coalescing pair $(i,j)$. For our purpose of
finding $R_{max}$, we just need to choose a specific pattern of nonzero
elements. To get that pattern we need to form a matrix that contains the
i-th column of the correct power of $AM$ as shown in Table VIII. This is an
example of the matrix that is used to determine the steps (distance) from
the first microdomain $i=1$. For example, for $j=8$, we need just the first
7 rows of the above matrix. The valid steps are the nonzero elements of the
matrix, and more specifically the various diagonal ones. The actual numbers
in the above matrix dont mean anything. What matters is whether they are
zero or not.

The program that calculates $R_{max}$, starts from the initial microdomain,
and goes to the neighboring one in a certain direction (left and right
diagonal, since we count the distance between neighbors). There are two
different directions that the steps (distance) can be taken, indicated by
the yellow color (from left to right), and green (from right to left). In
the Matlab program for this algorithm, this information is kept in a cell
array of the position in the matrix and the direction (left/right). An
additional feature is that when the step comes to a microdomains index that
is a part of a pairing (as microdomains 3-9 in the previous example) the
step splits into left and right. This is indicated by the light blue colored
elements in the matrix above. Associating the radii of the microdomains with
the elements in the above matrix, and the transitions from row to row in a
certain direction with the distance between the microdomains, gives an easy
way to find the distance $(R_{max})$ between any of the microdomains.
\medskip \bigskip

\qquad \pagebreak

SUMMARY

The diffusion-collision model was suggested in 1976 as a model for the
process of protein folding based on the dynamical interactions of
microdomains, as a series of diffusion-collision steps. Since then the
calculational power of computers has greatly increased, which enables faster
calculations in the model. However the combinatorial complexity for the
calculation of states, independent pathways, transitions, and $R_{max}$ is
too lengthy, and susceptible to error, if carried out by hand for each
protein to be studied. Introducing the two described algorithms reduces all
of these calculations to several minutes, even for a large number of
pairings. The algorithms have been implemented as MATLAB programs. For a
particular protein, the number of microdomains, the number of pairings and
actual pairs, microdomain radii and distances between microdomains are input
parameters.

\qquad \pagebreak

\pagebreak

$\stackrel{\stackrel{}{\text{\textsf{Table I. Example of a three microdomain
protein data.}}^{\text{\textsf{*}}}}}{\stackunder{^{\text{\textsf{*}}}\text{%
\textsf{Numerical values of these properties are the input parameters in
diffusion-collision model calculations.}}}{\pagebreak \stackrel{}{%
\stackunder{}{
\begin{tabular}{cccc}
\textsf{Microdomain} & \textsf{Radius (\AA )} & \textsf{Distance (\AA )} & 
\textsf{Access area (\AA 2)} \\ \hline\hline
\textsf{A} & \textsf{rA} &  & \textsf{A}$_A$ \\ \hline
\textsf{B} & \textsf{rB} &  & \textsf{A}$_B$ \\ \hline
\textsf{C} & \textsf{rC} &  & \textsf{A}$_C$ \\ \hline
\textsf{AB} &  & \textsf{cAB} & \textsf{A}$_{AB}$ \\ \hline
\textsf{AC} &  &  & \textsf{A}$_{AC}$ \\ \hline
\textsf{BC} &  & \textsf{cBC} & \textsf{A}$_{BC}$ \\ \hline
\textsf{ABC} &  &  & \textsf{A}$_{ABC}$ \\ \hline\hline
\end{tabular}
}}}}$

\smallskip \medskip

\begin{center}

$\medskip \stackunder{\text{\textsf{binary numbers represents a coalesced
pair. The decimal number is obtained by adding one to the decimal value of
the binary number.}}}{\stackunder{^{\text{\textsf{*}}}\text{\textsf{A
schematic description of the possible kinetic intermediate states. Decimal
numbers enumerate the states, while the binary digit 1 in the}}}{\stackunder{%
}{\medskip \stackrel{\stackrel{}{\text{\textsf{Table II. Possible states
for a three-microdomain protein.}}^{\text{\textsf{*}}}}}{\stackrel{}{
\begin{tabular}{cccccccccccccccc}
&  & \textsf{Unfolded} &  &  &  & \textsf{One-Pair} &  &  &  &
\textsf{Two-Pair} &  &  &  & \textsf{Folded} &  \\ 
\cline{3-3}\cline{7-7}\cline{11-11}\cline{15-15}
\textsf{\#} & \textsf{BC} & \textsf{AC} & \textsf{AB} & \textsf{\#} & 
\textsf{BC} & \textsf{AC} & \textsf{AB} & \textsf{\#} & \textsf{BC} & 
\textsf{AC} & \textsf{AB} & \textsf{\#} & \textsf{BC} & \textsf{AC} & 
\textsf{AB} \\ \hline\cline{2-4}\cline{6-8}\cline{10-12}\cline{14-16}
&  &  &  & \textsf{2} & \textsf{0} & \textsf{0} & \textsf{1} & \textsf{4} & 
\textsf{0} & \textsf{1} & \textsf{1} &  &  &  &  \\ \cline{6-8}\cline{10-12}
\textsf{1} & \textsf{0} & \textsf{0} & \textsf{0} & \textsf{3} & \textsf{0}
& \textsf{1} & \textsf{0} & \textsf{6} & \textsf{1} & \textsf{0} & \textsf{1}
& \textsf{8} & \textsf{1} & \textsf{1} & \textsf{1} \\ 
\cline{2-4}\cline{6-8}\cline{10-12}\cline{14-16}
&  &  &  & \textsf{5} & \textsf{1} & \textsf{0} & \textsf{0} & \textsf{7} & 
\textsf{1} & \textsf{1} & \textsf{0} &  &  &  &  \\ \cline{6-8}\cline{10-12}
\hline\hline\end{tabular}
}}}}}$
\end{center}

\qquad \pagebreak \medskip

\pagebreak \medskip

\smallskip

\qquad \pagebreak \medskip

\pagebreak

$\stackrel{\stackrel{}{\text{\textsf{Table III. Transition States, Bonds and
Parameters of three-microdomain protein.}}^{\text{\textsf{*}}}}}{\stackunder{%
\text{\textsf{and the folding and unfolding rates by solving the diffusion
equation.}}}{\stackunder{\text{\textsf{model calculations. The data is then
used to calculate the probabilities of the kinetic states }}}{\stackunder{^{%
\text{\textsf{*}}}\text{\textsf{Based on the initial parameters, the data in
this table is obtained from the diffusion-collision}}}{\stackrel{}{%
\stackunder{}{
\begin{tabular}{cccccl}
\textsf{Transition} & \textsf{Initial State} & \textsf{Bond Formed} & 
\textsf{Final State} & $\mathsf{R}_{min}$ & $\mathsf{R}_{max}$ \\ 
\hline\hline
\textsf{1-\TEXTsymbol{>}2} & \textsf{A-B-C} & \textsf{AB} & \textsf{AB-C} & 
\textsf{rA+rB} & \textsf{rA+cAB+rB} \\ 
\textsf{1-\TEXTsymbol{>}3} & \textsf{A-B-C} & \textsf{AC} & \textsf{B-AC} & 
\textsf{rA+rC} & \textsf{rA+cAB+2rB+cBC+rC} \\ 
\textsf{1-\TEXTsymbol{>}5} & \textsf{A-B-C} & \textsf{BC} & \textsf{A-BC} & 
\textsf{rB+rC} & \textsf{rB+cBC+rC} \\ \hline
\textsf{2-\TEXTsymbol{>}4} & \textsf{AB-C} & \textsf{AC} & \textsf{ABC}$_1$
& \textsf{rAB+rC} & \textsf{rAB+cBC+rC} \\ 
\textsf{2-\TEXTsymbol{>}6} & \textsf{AB-C} & \textsf{BC} & \textsf{ABC}$_2$
& \textsf{rAB+rC} & \textsf{rAB+cBC+rC} \\ 
\textsf{3-\TEXTsymbol{>}4} & \textsf{B-AC} & \textsf{AB} & \textsf{ABC}$_1$
& \textsf{rB+rAC} & \textsf{rB+cBC+rAC} \\ 
\textsf{3-\TEXTsymbol{>}7} & \textsf{B-AC} & \textsf{BC} & \textsf{ABC}$_3$
& \textsf{rB+rAC} & \textsf{rB+cBC+rAC} \\ 
\textsf{5-\TEXTsymbol{>}6} & \textsf{A-BC} & \textsf{AB} & \textsf{ABC}$_2$
& \textsf{rA+rBC} & \textsf{rA+cAB+rBC} \\ 
\textsf{5-\TEXTsymbol{>}7} & \textsf{A-BC} & \textsf{AC} & \textsf{ABC}$_3$
& \textsf{rA+rBC} & \textsf{rA+cAB+rBC} \\ \hline
\textsf{4-\TEXTsymbol{>}8} & \textsf{ABC}$_1$ & \textsf{BC} & \textsf{ABC}$%
_4 $ & \textsf{rB+rC} & $\pi $\textsf{rABC} \\ 
\textsf{6-\TEXTsymbol{>}8} & \textsf{ABC}$_2$ & \textsf{AC} & \textsf{ABC}$%
_4 $ & \textsf{rA+rC} & $\pi $\textsf{rABC} \\ 
\textsf{7-\TEXTsymbol{>}8} & \textsf{ABC}$_3$ & \textsf{AB} & \textsf{ABC}$%
_4 $ & \textsf{rA+rB} & $\pi $\textsf{rABC} \\ \hline\hline
\end{tabular}
}}}}}}$

\smallskip \medskip

$\stackrel{\stackrel{}{\text{\textsf{Table IV. Combinatorial
dependence on the number of pairings.}}^{\text{\textsf{*}}}}}{\stackrel{}{%
\stackunder{\text{\textsf{complexity in the calculations.}}}{\stackunder{%
\text{\textsf{of states, transitions, and independent pathways increases
quickly, thus creating combinatorial}}}{\stackunder{^{\text{\textsf{*}}}%
\text{\textsf{Considering large rnumber of pairings n, between the
microdomains of the protein, the number}}}{\stackunder{}{
\begin{tabular}{cccc}
\hline
\textsf{\# of pairings} & \textsf{\# of states} & \textsf{\# of independent }
& \textsf{\# of transitions} \\ 
\textsf{n} & \textsf{2}$^{\mathsf{n}}$ & \textsf{pathways n!} & \textsf{n2}$%
^{\mathsf{n-1}}$ \\ \hline\hline
\textsf{1} & \textsf{2} & \textsf{1} & \textsf{1} \\ 
\textsf{2} & \textsf{4} & \textsf{2} & \textsf{4} \\ 
\textsf{3} & \textsf{8} & \textsf{6} & \textsf{12} \\ 
\textsf{4} & \textsf{16} & \textsf{24} & \textsf{32} \\ 
\textsf{5} & \textsf{32} & \textsf{120} & \textsf{80} \\ 
\textsf{6} & \textsf{64} & \textsf{720} & \textsf{192} \\ 
\textsf{7} & \textsf{128} & \textsf{5040} & \textsf{448} \\ 
\textsf{8} & \textsf{256} & \textsf{40320} & \textsf{1024} \\ 
\textsf{9} & \textsf{512} & \textsf{362880} & \textsf{2304} \\ \hline
\end{tabular}
}}}}}}$

\qquad \pagebreak \medskip

\pagebreak \medskip

\smallskip

\qquad \pagebreak \medskip

\pagebreak  

$\stackrel{\stackrel{}{\text{\textsf{Table V. Simple cellular automaton
pattern}}}}{\stackunder{}{\stackunder{}{\stackrel{}{
\begin{tabular}{ccccccccccccccc}
\hline
\textsf{0} & \textsf{0} & \textsf{0} & \textsf{0} & \textsf{0} & \textsf{0}
& \textsf{0} & \textsf{1} & \textsf{0} & \textsf{0} & \textsf{0} & \textsf{0}
& \textsf{0} & \textsf{0} & \textsf{0} \\ \cline{8-8}
\textsf{0} & \textsf{0} & \textsf{0} & \textsf{0} & \textsf{0} & \textsf{0}
& \textsf{1} & \textsf{0} & \textsf{1} & \textsf{0} & \textsf{0} & \textsf{0}
& \textsf{0} & \textsf{0} & \textsf{0} \\ \cline{7-7}\cline{9-9}
\textsf{0} & \textsf{0} & \textsf{0} & \textsf{0} & \textsf{0} & \textsf{1}
& \textsf{0} & \textsf{0} & \textsf{0} & \textsf{1} & \textsf{0} & \textsf{0}
& \textsf{0} & \textsf{0} & \textsf{0} \\ \cline{6-6}\cline{10-10}
\textsf{0} & \textsf{0} & \textsf{0} & \textsf{0} & \textsf{1} & \textsf{0}
& \textsf{1} & \textsf{0} & \textsf{1} & \textsf{0} & \textsf{1} & \textsf{0}
& \textsf{0} & \textsf{0} & \textsf{0} \\ 
\cline{5-5}\cline{7-7}\cline{9-9}\cline{11-11}
\textsf{0} & \textsf{0} & \textsf{0} & \textsf{1} & \textsf{0} & \textsf{0}
& \textsf{0} & \textsf{0} & \textsf{0} & \textsf{0} & \textsf{0} & \textsf{1}
& \textsf{0} & \textsf{0} & \textsf{0} \\ \cline{4-4}\cline{12-12}
\textsf{0} & \textsf{0} & \textsf{1} & \textsf{0} & \textsf{1} & \textsf{0}
& \textsf{0} & \textsf{0} & \textsf{0} & \textsf{0} & \textsf{1} & \textsf{0}
& \textsf{1} & \textsf{0} & \textsf{0} \\ 
\cline{3-3}\cline{5-5}\cline{11-11}\cline{13-13}
\textsf{0} & \textsf{1} & \textsf{0} & \textsf{0} & \textsf{0} & \textsf{1}
& \textsf{0} & \textsf{0} & \textsf{0} & \textsf{1} & \textsf{0} & \textsf{0}
& \textsf{0} & \textsf{1} & \textsf{0} \\ 
\cline{2-2}\cline{6-6}\cline{10-10}\cline{14-14}
\textsf{1} & \textsf{0} & \textsf{1} & \textsf{0} & \textsf{1} & \textsf{0}
& \textsf{1} & \textsf{0} & \textsf{1} & \textsf{0} & \textsf{1} & \textsf{0}
& \textsf{1} & \textsf{0} & \textsf{1} \\ \hline
\end{tabular}}}}}$\smallskip 

\medskip \smallskip

$\stackrel{\stackrel{}{\text{\textsf{Table VI. Adjacency matrix for N=11
with no pairings}}}}{\stackunder{}{\stackunder{}{\stackrel{}{
\begin{tabular}{cccccccccccc}
\hline
\textsf{AM(1,:)} & \textsf{0} & \textsf{1} & \textsf{0} & \textsf{0} & 
\textsf{0} & \textsf{0} & \textsf{0} & \textsf{0} & \textsf{0} & \textsf{0}
& \textsf{0} \\ \cline{3-3}
\textsf{AM(2,:)} & \textsf{1} & \textsf{0} & \textsf{1} & \textsf{0} & 
\textsf{0} & \textsf{0} & \textsf{0} & \textsf{0} & \textsf{0} & \textsf{0}
& \textsf{0} \\ \cline{2-2}\cline{4-4}
\textsf{AM(3,:)} & \textsf{0} & \textsf{1} & \textsf{0} & \textsf{1} & 
\textsf{0} & \textsf{0} & \textsf{0} & \textsf{0} & \textsf{0} & \textsf{0}
& \textsf{0} \\ \cline{3-3}\cline{5-5}
\textsf{AM(4,:)} & \textsf{0} & \textsf{0} & \textsf{1} & \textsf{0} & 
\textsf{1} & \textsf{0} & \textsf{0} & \textsf{0} & \textsf{0} & \textsf{0}
& \textsf{0} \\ \cline{4-4}\cline{6-6}
\textsf{AM(5,:)} & \textsf{0} & \textsf{0} & \textsf{0} & \textsf{1} & 
\textsf{0} & \textsf{1} & \textsf{0} & \textsf{0} & \textsf{0} & \textsf{0}
& \textsf{0} \\ \cline{5-5}\cline{7-7}
\textsf{AM(6,:)} & \textsf{0} & \textsf{0} & \textsf{0} & \textsf{0} & 
\textsf{1} & \textsf{0} & \textsf{1} & \textsf{0} & \textsf{0} & \textsf{0}
& \textsf{0} \\ \cline{6-6}\cline{8-8}
\textsf{AM(7,:)} & \textsf{0} & \textsf{0} & \textsf{0} & \textsf{0} & 
\textsf{0} & \textsf{1} & \textsf{0} & \textsf{1} & \textsf{0} & \textsf{0}
& \textsf{0} \\ \cline{7-7}\cline{9-9}
\textsf{AM(8,:)} & \textsf{0} & \textsf{0} & \textsf{0} & \textsf{0} & 
\textsf{0} & \textsf{0} & \textsf{1} & \textsf{0} & \textsf{1} & \textsf{0}
& \textsf{0} \\ \cline{8-8}\cline{10-10}
\textsf{AM(9,:)} & \textsf{0} & \textsf{0} & \textsf{0} & \textsf{0} & 
\textsf{0} & \textsf{0} & \textsf{0} & \textsf{1} & \textsf{0} & \textsf{1}
& \textsf{0} \\ \cline{9-9}\cline{11-11}
\textsf{AM(10,:)} & \textsf{0} & \textsf{0} & \textsf{0} & \textsf{0} & 
\textsf{0} & \textsf{0} & \textsf{0} & \textsf{0} & \textsf{1} & \textsf{0}
& \textsf{1} \\ \cline{10-10}\cline{12-12}
\textsf{AM(11,:)} & \textsf{0} & \textsf{0} & \textsf{0} & \textsf{0} & 
\textsf{0} & \textsf{0} & \textsf{0} & \textsf{0} & \textsf{0} & \textsf{1}
& \textsf{0} \\ \hline\end{tabular}}}}}$\smallskip

\smallskip \medskip

$\stackrel{\stackrel{}{\text{\textsf{Table VII. Adjacency matrix for N=11
with microdomains 3-9 pairing}}}}{\stackunder{}{\stackunder{}{\stackrel{}{
\begin{tabular}{cccccccccccc}
\hline
\textsf{AM(1,:)} & \textsf{0} & \textsf{1} & \textsf{0} & \textsf{0} & 
\textsf{0} & \textsf{0} & \textsf{0} & \textsf{0} & \textsf{0} & \textsf{0}
& \textsf{0} \\ \cline{3-3}
\textsf{AM(2,:)} & \textsf{1} & \textsf{0} & \textsf{1} & \textsf{0} & 
\textsf{0} & \textsf{0} & \textsf{0} & \textsf{0} & \textsf{1} & \textsf{0}
& \textsf{0} \\ \cline{2-2}\cline{4-4}\cline{10-10}
\textsf{AM(3,:)} & \textsf{0} & \textsf{1} & \textsf{0} & \textsf{1} & 
\textsf{0} & \textsf{0} & \textsf{0} & \textsf{1} & \textsf{0} & \textsf{1}
& \textsf{0} \\ \cline{3-3}\cline{5-5}\cline{9-9}\cline{11-11}
\textsf{AM(4,:)} & \textsf{0} & \textsf{0} & \textsf{1} & \textsf{0} & 
\textsf{1} & \textsf{0} & \textsf{0} & \textsf{0} & \textsf{1} & \textsf{0}
& \textsf{0} \\ \cline{4-4}\cline{6-6}\cline{10-10}
\textsf{AM(5,:)} & \textsf{0} & \textsf{0} & \textsf{0} & \textsf{1} & 
\textsf{0} & \textsf{1} & \textsf{0} & \textsf{0} & \textsf{0} & \textsf{0}
& \textsf{0} \\ \cline{5-5}\cline{7-7}
\textsf{AM(6,:)} & \textsf{0} & \textsf{0} & \textsf{0} & \textsf{0} & 
\textsf{1} & \textsf{0} & \textsf{1} & \textsf{0} & \textsf{0} & \textsf{0}
& \textsf{0} \\ \cline{6-6}\cline{8-8}
\textsf{AM(7,:)} & \textsf{0} & \textsf{0} & \textsf{0} & \textsf{0} & 
\textsf{0} & \textsf{1} & \textsf{0} & \textsf{1} & \textsf{0} & \textsf{0}
& \textsf{0} \\ \cline{7-7}\cline{9-9}
\textsf{AM(8,:)} & \textsf{0} & \textsf{0} & \textsf{1} & \textsf{0} & 
\textsf{0} & \textsf{0} & \textsf{1} & \textsf{0} & \textsf{1} & \textsf{0}
& \textsf{0} \\ \cline{4-4}\cline{8-8}\cline{10-10}
\textsf{AM(9,:)} & \textsf{0} & \textsf{1} & \textsf{0} & \textsf{1} & 
\textsf{0} & \textsf{0} & \textsf{0} & \textsf{1} & \textsf{0} & \textsf{1}
& \textsf{0} \\ \cline{3-3}\cline{5-5}\cline{9-9}\cline{11-11}
\textsf{AM(10,:)} & \textsf{0} & \textsf{0} & \textsf{1} & \textsf{0} & 
\textsf{0} & \textsf{0} & \textsf{0} & \textsf{0} & \textsf{1} & \textsf{0}
& \textsf{1} \\ \cline{4-4}\cline{10-10}\cline{12-12}
\textsf{AM(11,:)} & \textsf{0} & \textsf{0} & \textsf{0} & \textsf{0} & 
\textsf{0} & \textsf{0} & \textsf{0} & \textsf{0} & \textsf{0} & \textsf{1}
& \textsf{0} \\ \hline
\end{tabular}}}}}$

\qquad \pagebreak \medskip

\pagebreak \medskip

\smallskip

\qquad \pagebreak \medskip

\pagebreak  

$\stackrel{\stackrel{}{\text{\textsf{Table VIII. Adjacency matrix pattern
used to find R}}_{\text{\textsf{max}}}\text{.}}}{\stackunder{}{\stackunder{}{%
\stackrel{}{
\begin{tabular}{cccccccccccc}
\hline
\textsf{Initial} & \textsf{1} &  &  &  &  &  &  &  &  &  &  \\ \hline
\textsf{AM(1,:)} & \textsf{0} & \textsf{1} & \textsf{0} & \textsf{0} & 
\textsf{0} & \textsf{0} & \textsf{0} & \textsf{0} & \textsf{0} & \textsf{0}
& \textsf{0} \\ \cline{3-3}
\textsf{AM}$^{\mathsf{2}}$\textsf{(2,:)} & \textsf{1} & \textsf{0} & \textsf{%
1} & \textsf{0} & \textsf{0} & \textsf{0} & \textsf{0} & \textsf{0} & 
\textsf{1} & \textsf{0} & \textsf{0} \\ \cline{4-4}\cline{10-10}
\textsf{AM}$^{\mathsf{3}}$\textsf{(3,:)} & \textsf{0} & \textsf{3} & \textsf{%
0} & \textsf{2} & \textsf{0} & \textsf{0} & \textsf{0} & \textsf{2} & 
\textsf{0} & \textsf{2} & \textsf{0} \\ 
\cline{3-3}\cline{5-5}\cline{9-9}\cline{11-11}
\textsf{AM}$^{\mathsf{4}}$\textsf{(4,:)} & \textsf{3} & \textsf{0} & \textsf{%
9} & \textsf{0} & \textsf{2} & \textsf{0} & \textsf{2} & \textsf{0} & 
\textsf{9} & \textsf{0} & \textsf{2} \\ 
\cline{2-2}\cline{6-6}\cline{8-8}\cline{10-10}\cline{12-12}
\textsf{AM}$^{\mathsf{5}}$\textsf{(5,:)} & \textsf{0} & \textsf{21} & 
\textsf{0} & \textsf{20} & \textsf{0} & \textsf{4} & \textsf{0} & 2\textsf{0}
& \textsf{0} & 2\textsf{0} & \textsf{0} \\ \cline{7-7}
\textsf{AM}$^{\mathsf{6}}$\textsf{(6,:)} & \textsf{21} & \textsf{0} & 81 & 
\textsf{0} & \textsf{24} & \textsf{0} & \textsf{24} & \textsf{0} & \textsf{81%
} & \textsf{0} & 2\textsf{0} \\ \cline{6-6}\cline{8-8}
\textsf{AM}$^{\mathsf{7}}$\textsf{(7,:)} & \textsf{0} & \textsf{183} & 
\textsf{0} & \textsf{186} & \textsf{0} & \textsf{48} & \textsf{0} & \textsf{%
186} & \textsf{0} & \textsf{182} & \textsf{0} \\ \cline{5-5}\cline{9-9}
\textsf{AM}$^{\mathsf{8}}$\textsf{(8,:)} & \textsf{183} & \textsf{0} & 
\textsf{737} & \textsf{0} & \textsf{234} & \textsf{0} & \textsf{234} & 
\textsf{0} & \textsf{737} & \textsf{0} & \textsf{182} \\ 
\cline{4-4}\cline{10-10}
\textsf{AM}$^{\mathsf{9}}$\textsf{(9,:)} & \textsf{0} & \textsf{1657} & 
\textsf{0} & \textsf{1708} & \textsf{0} & \textsf{468} & \textsf{0} & 
\textsf{1708} & \textsf{0} & \textsf{1656} & \textsf{0} \\ 
\cline{3-3}\cline{5-5}\cline{9-9}\cline{11-11}
\textsf{AM}$^{\mathsf{10}}$\textsf{(10,:)} & \textsf{1657} & \textsf{0} & 
\textsf{6729} & \textsf{0} & \textsf{2176} & \textsf{0} & \textsf{2176} & 
\textsf{0} & \textsf{6729} & \textsf{0} & \textsf{1656} \\ 
\cline{2-2}\cline{6-6}\cline{8-8}\cline{12-12}
\textsf{AM}$^{\mathsf{11}}$\textsf{(11,:)} & \textsf{0} & \textsf{15115} & 
\textsf{0} & \textsf{15634} & \textsf{0} & \textsf{4352} & \textsf{0} & 
\textsf{15634} & \textsf{0} & \textsf{15114} & \textsf{0} \\ \hline
\end{tabular}
}}}}$

\qquad \pagebreak \medskip

\pagebreak \medskip

\smallskip

\qquad \pagebreak \medskip

\pagebreak  

$\stackunder{\text{\textsf{There are 8
states, 6 ways to get from state 1 to state 8 and 12 transitions
represented
by arrows.}}}{\stackunder{\mathsf{A\ schematic\ view\ of\ the\ states\
and\ transitions\ for\ 3\ microdomains\ with\
n=3.}}{\stackrel{\stackrel{}{\text{%
\textsf{Figure 2.}}}}{\stackrel{}{\stackunder{}{\FRAME{itbpF}{190.75pt}{%
90.375pt}{0pt}{}{}{Figure }{\includegraphics{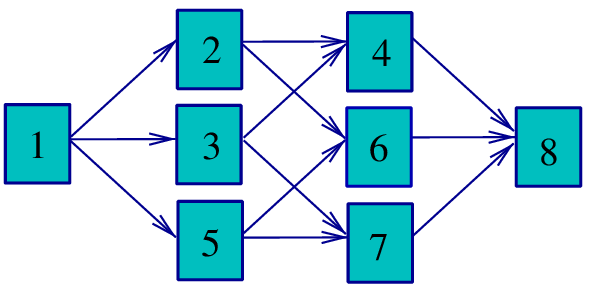}}}}}}}$

\qquad \pagebreak \medskip

\pagebreak \medskip

\smallskip

\qquad \pagebreak \medskip

\pagebreak

$\stackunder{\text{\textsf{A representation of an 11 - microdomain unfolded
protein chain.}}}{\stackrel{\stackrel{}{\text{\textsf{Figure 3.}}}}{%
\stackrel{}{\stackunder{}{\FRAME{itbpF}{366.75pt}{65.6875pt}{0pt}{}{}{Figure 
}{\includegraphics{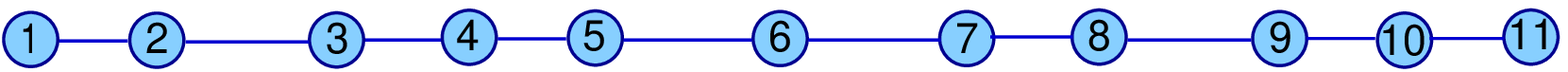}}}}}}$

\qquad \pagebreak \medskip

\pagebreak \medskip

\smallskip

\qquad \pagebreak \medskip

\pagebreak

$\stackunder{\text{\textsf{The pairing has changed the adjacency matrix.}}}{%
\stackunder{\text{\textsf{A representation of an 11 - microdomain protein
chain with one coalesced pairing between microdomains 3 and 9.}}}{\stackrel{%
\stackrel{}{\text{\textsf{Figure 4.}}}}{\stackrel{}{\stackunder{}{\FRAME{%
itbpF}{295.625pt}{146pt}{0pt}{}{}{Figure }
{\includegraphics{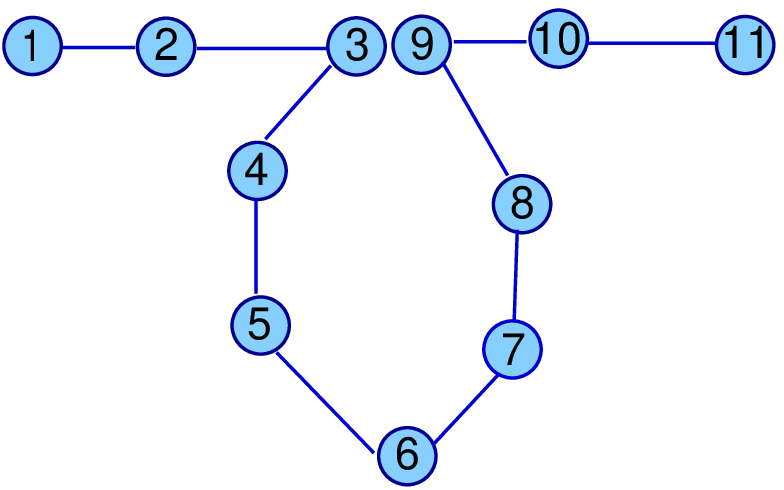}}}}}}}$

\end{document}